\documentclass{mem}
\usepackage{natbib}
\usepackage{graphicx}
\usepackage[a4paper]{hyperref}
\idline{00}{001}
\begin{document}

\title{LISA: Mars and the limits of life}

   \subtitle{}

\author{
G. \,Galletta\inst{1,2}
\and
M. D'Alessandro\inst{3}
\and
G. Bertoloni\inst{4}
\and
G. Fanti\inst{2,5}
\and
E. Dainese\inst{4}
\and
M. Pelizzo\inst{6}
\and
F. Ferri\inst{2}
\and
D. Pavarin\inst{2,5}
\and
C. Bettanini\inst{2}
\and
G. Bianchini\inst{2,5}
\and
S. Debei\inst{2,5}
          }

\offprints{G. Galletta}

\institute{Dipartimento di Astronomia, vicolo Osservatorio, 3, 35122 Padova, Italy
\and
CISAS "G. Colombo", Via Venezia 15, 35131 Padova, Italy
\and
INAF, Osservatorio Astronomico di Padova, vicolo Osservatorio 5, 35122 Padova, Italy
\and
Dipartimento di Istologia, Microbiologia e Biotecnologie Mediche, via Gabelli 63, 35122 Padova, Italy
\and
Dipartimento di Ingegneria Meccanica, Via Venezia, 1, 35131 Padova, Italy
\and
CNR-INFM LUXOR Laboratory, Via Gradenigo, 6B - 35131 Padova, Italy
\email{giuseppe.galletta@unipd.it}
}

\authorrunning{Galletta et al.}

\titlerunning{LISA: Mars simulator}

\abstract{We describe the results of the first tests made on LISA, a simulator of planetary 
environments designed and built in Padua, dedicated to the study of the limit of bacterial 
life on the planet Mars. Tests on the cryogenic circuit, on the UV illumination and on 
bacterial coltures at room temperature that shall be used as references are described.
}
\maketitle{}

\section{Introduction}

During 2004 and 2005, financed by the university of Padua, we designed and built a
simulator of planetary environments (LISA=Laboratorio Italiano di Simulazione Ambienti).
This is the first, and at present the only, environmental simulator of its kind in Italy.
Its versatility makes it able to reproduce a wide range of controlled and stable situations, from
``warm" (human body or terrestrial conditions) to ``cold" (from Antarctica to Mars) with six different
samples that may be exposed to the same environment simultaneously. A full description of the
simulator has been presented by \citet{galletta}.

The simulator is currently used as a research tool for Astrobiology or Planetary
Geology in extreme conditions of temperature, pressure and atmosphere, not achievable in other ways
in terrestrial laboratories. In particular, we want to reproduce the Martian environments, with a
group of biologists interested in the study of bacterial metabolism in harsh environments.

The planet Mars is a privileged place in the Solar System on which to search for lifeforms, past or present.
In fact, for several years it has been known that Mars had water on its surface\citep{Nyquist,Head2}. 
In 1976 the Viking 1 and 2 probes landed on Mars and performed
experiments aimed at the detection of life. They produced data whose interpretation has been doubtful
or ambiguous \citep[see][]{Burgess}. Subsequently, the analysis of meteorites such as ALH 84001 \citep{Mckay}
with possible traces of fossil bacteria, have reopened the debate.

While waiting for new experiments, it is necessary to understand the possibilities of finding life in
the present Martian conditions. The surface is not protected like that of the Earth by an atmosphere
with an ozone layer: the ground pressure is 6 mbar and the daily UV flux is 361 kJ/m$^2$, compared to 1000
mbar and 39 kJ/m$^2$ on the Earth \citep{Cockell}. The consequent lack of greenhouse effect
produces a mean temperature between -70 and -10$^\circ$C at a latitude of 20$^\circ$ north (the landing site 
of Viking 1 and Pathfinder) depending on the Sun's elevation above the horizon.
From radiometric data of the probes orbiting around Mars, the temperature can locally reach
27$^\circ$C while at the polar caps in the winter can fall to -143$^\circ$C. These cold temperatures
and these excursions are impossible on the Earth and must be simulated in the laboratory.

If life was born on Mars billions of years ago, it may have found suitable conditions beneath the surface, where
temperature and pressure increase with depth. A stable environment may exist under the duricrust or
the dust cover, with annual thermal stability ($<10^\circ$C) at a depth of 10 cm, with temperatures
between -70 e -40$^\circ$C and shielded from UV rays. Water may be liquid at depths betweeen 150 m (0,7 atm) 
and 8 km (600 atm), according to the different models of the Martian surface. Terrestrial lifeforms exist
(barophiles) that are able to live at those pressures. Occasionally, the arrival on the surface of some bacteria
can determine the extinction or mutations of more UV resistant species.

Before reaching the red planet, we may explore more economically what kind of lifeforms may live
in such an extreme environment. LISA satisfies all these conditions because it is a relatively easy way to
study what happens in the environmental conditions on the planet Mars. In doing the biological experiments, we must take
into account some caveats: 1) life on a planet different from that on Earth may use a completely different
combination of nucleic acids and amino acids with respect to the first terrestrial lifeforms; 2) we
don't have (yet) either Martian bacteria or Martian soils for the test. We use terrestrial
surrogates!

\section{Tests of experimental conditions}

The sample to be studied (ground, chemical substances or bacterial cultures) is deposited in
containers similar to Petri dishes, but specifically produced in aluminium by the
mechanical laboratories of the Asiago Astronomical Observatory, that can be cooled or warmed by
contact. In order to guarantee experimental sterility, the culture containers are hermetically
closed in steel vessels (reaction cells) of about 250 cc. These vessels are connected with the exterior via a pipe 
equipped with mechanical filters of suitable porosity (based on the
dimensions of the biological samples used). These filters allow the passage of gases through
the pumps but prevent the spread of the micro-organisms or particles in the cryostat chamber.

The vessels containing the biological samples are cooled by contact with a large aluminium
dish that is at the top of a small reservoir of about 2000 cc, filled from a larger reservoir of
Ranger Air Liquide at a pressure of 2 bar. A PT100 temperature sensor connected with an electric
valve opens or close the liquid nitrogen flux of the Ranger reservoir. If needed, an electric resistance
between the dish and the reservoir is open, in a feedback circuit that keeps the vessels
temperature fixed within $\pm0.5^\circ$.

\begin{figure}
\resizebox{\hsize}{!}{\includegraphics[clip=true]{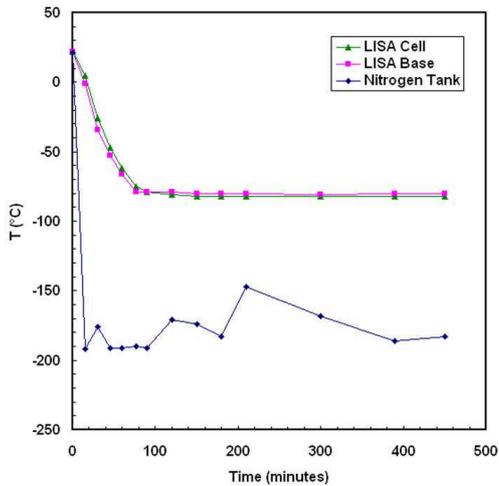}}
\caption{\footnotesize
Decrease of Temperature in the different portions of the Simulator }
\label{T}
\end{figure}

The thermal efficiency of LISA was studied by cooling the simulation chamber with a large flux of liquid
nitrogen and measuring the inner temperatures of the chamber and of the vessels at different points.
We measured with several PT100 sensors the temperature of the reservoir, the dish and the vessels,
at the onset and the end of the flux of liquid nitrogen, calculating how fast the temperature changes in the
different parts of LISA. The results are shown in Figure \ref{T}. The goal temperature selected was -80
$^\circ$C, maintained by the feedback circuit(valve+electric resistance). The target temperature has been
reached by the floor of the vessels in $\sim$80 min, starting from a room temperature of 25 $^\circ$C.
The calculated cooling rate of the dish and vessels is not linear and is approximated by a gradient
of -0.78$\pm$0.06 $^\circ$/min. When swithched off, the system
(dish plus vessels) keeps its temperature within 1$^\circ$C for about 20 minutes, starting to
increase thereafter.

\begin{figure}
\resizebox{\hsize}{!}{\includegraphics{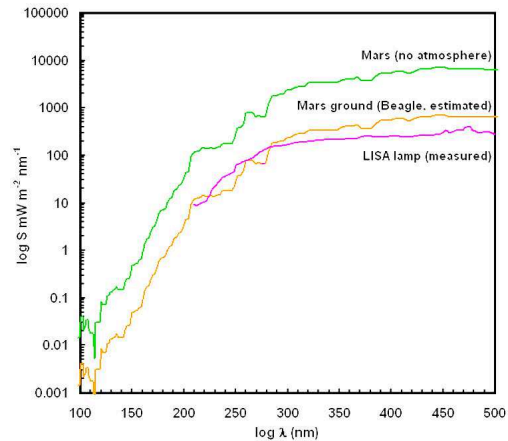}}
\caption{\footnotesize Flux of LISA lamp compared with that of Mars (outside the atmosphere and
estimated on the ground).}
\label{UV}
\end{figure}

In the upper part of the vessels quartz windows are located, allowing the passage of
radiation, including UV rays longer than 160 nm. Different types and thicknesses of UV filters 
may be placed over the windows to simultaneously evaluate the sample response to various doses of
radiation.

To test the performances of the UV illumination system, we estimated the UV flux received by
the samples using a photo-spectrometer with a sensor, placed on the dish at the location of the
vessels. It registered how much UV was falling on the vessels after passing through
the quartz windows. The values were taken in different positions and calibrated
using various UV filters with passbands of 20 $\mu$m. The measured flux is shown in Figure
\ref{UV} and compared with the solar flux received by Mars outside the atmosphere and estimated
at ground level \citep{Cockell}.

The Martian atmosphere is composed mainly of CO$_2$, with a small percentage of N$_2$ (see Table
\ref{atm}). Its complete reproduction is very hard, and we used a mixture of gases ordered from a 
specialised factory, whose composition is shown in the third column of the Table. This mixture
is quite expensive, so for the majority of the tests described here we used simply CO$_2$ gas.

\begin{table}
\caption{Martian and LISA atmosphere}
\label{atm}
\begin{center}
\begin{tabular}{lcc}
\hline
\\
Gas & Mars & LISA \\
\\
\hline
\\
CO$_2$ & 95.32\%	& 95.5\% \\
N$_2$  & 2.7\%	& 3\% \\
Ar 	 & 1.6-1,7\% & 1.6\% \\
O$_2$  & 0.13-0.2\% & 0.13\% \\
CO 	& 0.08\% & 0.07\% \\
H$_2$O & 210 ppm & absent \\
NO 	& 100 ppm & absent \\
Ne  & 2.5 ppm & 300 ppm \\
HDO &	0.85 ppm & absent \\
Kr &	0.3 ppm  & 300 ppm \\
Xe & 0.08 ppm & 300 ppm \\
\\
\hline
\end{tabular}
\end{center}
\end{table}

\begin{figure}
\resizebox{\hsize}{!}{\includegraphics[clip=true]{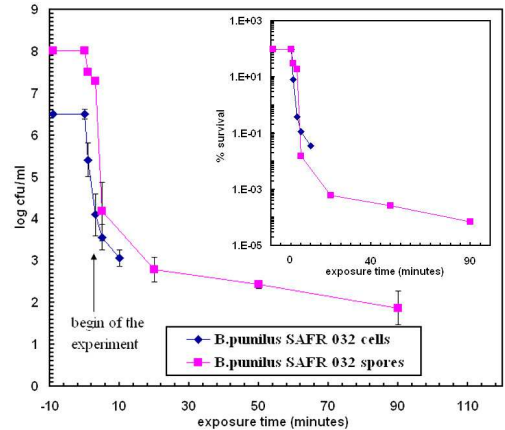}}
\caption{\footnotesize Survival curves of {\it Bacillus pumilus} (spores and vegetative cells) 
at room temperature (25$^\circ$C), 7 mbar of CO$_2$, vs. time, irradiated with the lamp whose 
flux is shown in Figure \ref{UV}. In the inset, the percentage of survival is shown.}
\label{bio}
\end{figure}

To test the procedure for biological samples, the biologists of our team cultured different bacterial
strains in their laboratories. These were then placed in LISA and irradiated by the UV lamp with
different exposure times by initially covering all the quartz windows with
aluminium diaphragms, and then opening them in nested intervals. At the end of the experiments, all
the six bacterial samples were kept at the same atmospheric pressure and temperature conditions but
received different doses of radiation. Then the pressure and eventually the temperature were brought
to normal conditions and the bacterial sample, extracted from the vessels, analysed in
order to estimate the degree of deactivation or the metabolic variations.
The first experiments performed are ``by reference" at room temperature. 
The survival curves exposed in Figure \ref{bio} demonstrate the higher survival capability of 
spores than vegetative cells obtained from the same bacterial strain. One of them is shown in 
Figure \ref{bio}. We used cultures of {\it Deinococcus radiodurans}, particularly resistant to ionizing 
radiations such as gamma rays, {\it Bacillus nealsonii} and {\it Bacillus pumilus}.

\section{Conclusions}

The LISA simulator has been tested with the conditions of the planet Mars and is now starting
biological experiments to understand what kind of (terrestrial) lifeforms may survive on the
surface or sub-surface. LISA may also be used for the testing of materials
and samples in conditions similar to those of Antarctica or colder environments. These tests have 
shown that LISA is able to simulate the environmental conditions on Mars at various latitudes.

\begin{acknowledgements}
The authors thanks D. Garoli, V. Da Deppo and P. Nicolosi for their help in measuring the
absolute flux of UV. This research is supported by the University of Padua funds (ex 60\%)
2006. The liquid nitrogen supply is kindly furnished by Air Liquide Italy - North-East region.
We also thank the factory CINEL of Vigonza-Padova for their technical support and all the 
mechanical and electronic staff of Ekar and Padua Observatory.

\end{acknowledgements}

\bibliographystyle{aa}

\end{document}